\def\spose#1{\hbox to 0pt{#1\hss}}
\def\approxlt{\mathrel{\spose{\lower 3pt\hbox{$\sim$}}
	\raise 2.0pt\hbox{$$<$$}}}
\def\approxgt{\mathrel{\spose{\lower 3pt\hbox{$\sim$}}
	\raise 2.0pt\hbox{$>$}}}

\def\multleft#1{\hbox to size{\vbox {\halign {\lft{##}\cr #1}}\hfill}\par}
\def\multright#1{\hbox to size{\vbox {\halign {\rt{##}\cr #1}}\hfill}\par}

\def\today{\ifcase\month\or January\or February\or March\or April\or May\or
      June\or July\or August\or September\or October\or November\or December\fi
      \space\number\day, \number\year}
\def\$<${\thinspace}
\def\s{\hbox{\phantom{5}}}	

\def\boxit#1{\vbox{\hrule\hbox{\vrule\kern3pt\vbox{\kern3pt
          #1 \kern3pt}\kern3pt\vrule}\hrule}}

\def\cm{{\rm\thinspace cm}}

\def\erg{{\rm\thinspace erg}}

\def\K{{\rm\thinspace K}}
\def\keV{{\rm\thinspace keV}}
\def\km{{\rm\thinspace km}}
\def\kpc{{\rm\thinspace kpc}}

\def\Mpc{{\rm\thinspace Mpc}}
\def\Msun{\hbox{$\rm\thinspace M_{\odot}$}}

\def\s{{\rm\thinspace s}}
\def\yr{{\rm\thinspace yr}}

\def\ergpcmsqps{\hbox{$\erg\cm^{-2}\s^{-1}\,$}}
\def\ergpcmsqparcsecps{\hbox{$\erg\cm^{-2}\rm{arcsec}^{-2}\s^{-1}\,$}}

\def\ergps{\hbox{$\erg\s^{-1}\,$}}

\def\kmps{\hbox{$\km\s^{-1}\,$}}

\def\Msunpyr{\hbox{$\Msun\yr^{-1}\,$}}

\def\pcmK{\hbox{$\cm^{-3}\K$}}

\def\kmpspMpc{\hbox{$\kmps\Mpc^{-1}$}}

\def\othree{[OIII]$\lambda5007$}
\def\otwo{[OII]$\lambda3727$}
\def\Lya{Ly$\alpha$}

\documentstyle[psfig]{mn}
\begin{document}
\hsize=6truein

\title{Extended line emission around seven radio-loud quasars at redshift $z \sim 2$}

\author[]
{\parbox[]{6.in} {R.J.~Wilman, R.M.~Johnstone and C.S.~Crawford \\
\footnotesize
Institute of Astronomy, Madingley Road, Cambridge CB3 0HA \\ }}
\maketitle

\begin{abstract}
We present near-infrared spectra of seven radio-loud quasars with a median redshift of 2.1, five of which were previously known to have \Lya~nebulae. Extended [OIII]$\lambda5007$ and H$\alpha$ emission are evident around six objects, at the level of a few times $10^{-16}$\ergpcmsqparcsecps~within $\simeq 2$ arcsec of the nucleus ($\equiv 16$\kpc~in the adopted cosmology). Nuclear [OII]$\lambda3727$ is detected in three of the 5 quasars studied at this wavelength, and clearly extended in one of them.  

The extended [OIII] tends to be brighter on the side of the nucleus with the stronger, jet-like radio emission, indicating at least that the extranuclear gas is distributed anisotropically. It is also typically redshifted by several hundred \kmps~from the nuclear [OIII], perhaps due to the latter being blueshifted from the host galaxy's systemic velocity. Alternatively, the velocity shifts could be due to infall (which is suggested by line widths $\sim1000$\kmps~FWHM) in combination with a suitable dust geometry. \Lya/H$\alpha$ ratios well below the case B value suggest that some dust is present.

Photoionization modelling of the [OIII]/[OII] ratios in the extended gas suggests that its pressure is around or less than a few times $10^{7}$\pcmK; any confining intracluster medium is thus likely to host a strong cooling flow. A comparison with lower-redshift work suggests that there has been little evolution in the nuclear emission line properties of radio-loud quasars between redshifts 1 and 2. 
\end{abstract}

\section{INTRODUCTION}
The processes which precipitated the steep decline in the co-moving density of luminous quasars since the epoch $z\sim 2$ remain poorly understood. One observational approach to the problem proceeds through the study of the emission-line gas extending tens of kiloparsecs from the quasar. This material is the only visible tracer of the quasar's immediate environment, and as such its kinematics and ionization state may yield clues to the mechanisms which have driven the cosmological evolution of the population as a whole, as well as the early stages of galaxy formation more generally.

There are numerous indications that at redshift $z>0.5$ powerful radio sources inhabit environments similar to the rich clusters found locally around the luminous FR~II sources Cygnus A and 3C295. The radio source properties alone demand a certain type of environment: a confining medium capable of acting as a working surface for the formation of steep-spectrum radio lobes is a basic requirement. Faraday depolarization asymmetry (Garrington \& Conway~1991), distorted/compressed radio source morphologies at high redshift (Hintzen et al.~1983; Barthel \& Miley 1988) and the discovery of sources with very high Faraday rotation measures (Carilli et al.~1994, 1997) suggest that the medium is clumpy and dense. Of particular interest is the so-called `alignment effect', whereby the radio and extended optical/UV emission of $z>0.6$ radio galaxies are seen to be coincident (McCarthy et al.~1987; Chambers et al.~1987); recent work by Lacy et al.~(1999) suggests that different physical mechanisms are responsible for its manifestation on different spatial scales. Similarly, Best et al.~(2000a,b) have studied the emission-line properties of a large sample of $z\sim1$ radio galaxies: they conclude that in small radio sources ($\leq 150$\kpc~in size) the kinematics and ionization state of the line-emitting clouds are dominated by the effects of the bow shock associated with the expansion of the radio source through a confining medium; photoionization of the gas by the nuclear source is more important in larger radio sources.

To study the environments of radio-loud quasars in corresponding detail is technically more difficult, as the bright unresolved nucleus tends to overwhelm surrounding structure. Over the last decade or more, however, we have pioneered the use of optical long-slit spectroscopy to probe the extended emission-line regions around radio-loud quasars out to redshift $z\sim1$ (Fabian et al.~1988; Crawford \& Fabian 1989; Forbes et al.~1990; Bremer et al.~1992). Where the emission line ratio [OIII]$\lambda5007$/[OII]$\lambda3727$ can be measured for the extended gas, it can be used in conjunction with photoionization models to deduce the gas pressure. The latter is found to be consistently high ($nT \sim 5 \times 10^{5} - 1 \times 10^{7}$ \pcmK~at tens of kiloparsecs from the nucleus), and an increasing function of redshift (or of quasar luminosity). Pressure equilibrium between the line-emitting gas and any confining X-ray halo restricts the cooling time of the hot gas to be less than the Hubble time, suggesting that the quasars reside in cooling flows with mass deposition rates of up to $1000$\Msunpyr within 20\kpc. Such flows would have strongly influenced the formation and evolution of both the quasar and its host galaxy (see eg. Fabian \& Crawford~1990). This interpretation is supported by the detection of extended X-ray emission around 3CR quasars by Crawford et al.~(1999) and by Hardcastle \& Worrall~(1999).

Beyond $z \sim 1$ the diagnostic lines of [OIII] and [OII] move into the near-IR where differences in the technology and atmospheric properties conspire to make the extension of this technique more difficult. Fortunately, however, there is a narrow redshift window around $z=2.2$ where the [OII], [OIII]+H$\beta$ and H$\alpha$+[NII] complexes fall in the J, H and K atmospheric transmission bands, respectively. We have thus obtained near-IR spectra of 7 radio-loud quasars near this redshift using the CGS4 spectrograph on the United Kingdom Infrared Telescope (UKIRT). Studies from the ground (Heckman et al.~1991a,b, hereafter H91a,b) and with HST (eg. Lehnert et al.~1999a, hereafter L99a) have established that 100\kpc-scale \Lya~nebulae are common around high-redshift quasars, and a weaker analogue of the radio-galaxy alignment effect has been found (Lehnert et al.~1999b). Nevertheless, the spectroscopic elucidation of their properties remains in its infancy (eg. Lehnert \& Becker~1998). 

The structure of the paper is as follows: in section 2 we describe the observations and data reduction, and in section 3 detail the extraction of the extended emission and present the results on an object-by-object basis; in section 4 we
discuss its kinematics, morphology, ionization state and relationship with the radio source; section 5 comprises a brief analysis of the nuclear emission-line spectra in the light of lower redshift studies, and section~6 a summary of our findings and the conclusions. The cosmological parameters $H_{0}=50$\kmpspMpc, $q_{0}=0.5$ are adopted throughout, giving a spatial scale of 8\kpc~arcsec$^{-1}$ at $z=2.2$.

\begin{table*}
\begin{center}

\caption{Target properties and observation log}
\begin{tabular}{||lllllllll||} \hline
Object & Other name & z & spatial scale & Band$^{\dagger}$ & Exposure & Slit PA  &  Seeing$^{\star}$ & Comments \\  
       &            &   &  (kpc/arcsec) &   & (minutes)      & (deg. E of N)       & (arcsec) & \\ \hline 
\multicolumn{9}{c}{1997 November 23--24} \\ 
Q0017+154 & 3C9   & 2.012 & 8.2 & H+K & 116 & 0 & 1.4  & \\ 
          &       &       &     & J  & 90 & 0  & 1.3  & \\
Q0225-014 & PKS   & 2.037 & 8.2 & H+K & 60 & 0 & 1.4  & \\
Q0445+097 & PKS   & 2.11  & 8.1 & H+K & 60 & 0 & 1.3  & \\
          &       &       &     & J  & 45 & 0 & 1.5 & \\	
Q0802+103 & 3C191 & 1.956 & 8.2 & H+K & 46 & 0  & 1.3 & \\ 
	  &       &       &     &J  & 110 & 0   & 1.3 &\\ \\
\multicolumn{9}{c}{1999 June 26--27} \\ 
Q1658+575 & 4C57.29 & 2.173 & 8.0 & H+K & 52 & 45  & 1.5 & \\ 
 	  &         &       &     & H+K & 28 & 135 & 1.1 & $\ddagger$ with 1-pix wide slit \\
Q1816+475 & 4C47.48 & 2.225 & 8.0 & H+K & 60 & 146 & 1.0 & \\
	  &         &       &     & H+K & 56 & 56  & 1.2 & \\
	  &         &       &     & H+K & 56 & 45  & 1.1 & \\
          &         &       &     & J  & 64 & 45  & 1.2  & \\
Q2338+042 & PKS,4C04.81 & 2.594 & 7.6 & H+K & 36  & 90 & 0.9 & \\ 
          &             &       &     & H+K & 54 & 135 & 0.9 & \\
          &             &       &     &  J  & 108   & 135 & 1.0 & \\ \hline
\end{tabular}
\end{center}
$\dagger$ H+K covers [OIII]$\lambda5007,4959$ and [NII]+H$\alpha$; J covers [OII]$\lambda3727$\\
$\star$ FWHM of standard star spatial profile along the slit \\
$\ddagger$ at start of first night; slit-width changed thereafter to 2-pix (1.2 arcsec) for rest of run \\
\end{table*}

\section{OBSERVATIONS AND DATA REDUCTION}
\subsection{Target selection}
In order to maximise the likelihood of detecting extended emission, the targets were chosen primarily from those with known Ly$\alpha$ nebulae in the narrow-band imaging study of quasars in H91a. For other objects found in a NED search covering the appropriate area of the sky and redshift but not observed by H91a, the presence of strong, narrow nuclear HeII$\lambda1640$ was used as a further selection criterion. H91b confirmed the suggestion of Foltz et al.(1988) that this line plausibly originates in the inner parts of such nebulae and that it is thus an indirect indicator of their presence. Other desirable attributes for potential targets were a resolved radio structure (especially one with a depolarization asymmetry) and a well-constrained X-ray--optical spectral energy distribution (SED) for use in photoionization modelling. The sample is thus not in any sense statistically complete.

\subsection{Data acquisition and reduction}
The observations were made using the CGS4 spectrograph on UKIRT on the nights of 1997 November 23-24 and 1999 June 26-27. The 256x256 InSb array, 40 l/mm grating and 300mm focal length camera were in place, yielding a spatial scale of 0.61 arcsec per pixel and a spectral resolution of $\sim 850$\kmps FWHM (in the H-band) with a 2-pixel-wide slit. 

For each object we acquired a spectrum covering the [OIII]$\lambda 5007,4959$ doublet and the [NII]+H$\alpha$ complex (ie. connecting the H and K-bands) at at least one position angle, chosen to align with either the axis of greatest extent in published \Lya~maps or along/perpendicular to any extended radio structure. When time permitted, a J-band spectrum covering redshifted [OII]$\lambda3727$ was also obtained. The ND-STARE mode was used along with the conventional object-sky-sky-object nodding pattern, thus obviating the need for separate bias and dark current frames and permitting the computation of an external error for each pixel; it also reduces the required accuracy of the flat-field frame. Atmospheric absorption features were removed by ratioing with some of the main sequence F stars tabulated at the telescope which were, in most cases, photometrically calibrated using observations of faint standards. 

On- and off-line data reduction was performed using version V1.3-0 of the Portable CGS4 Data Reduction package available through Starlink. Row-by-row spectra were extracted from the fully-reduced spectral images and converted to ascii format for use with the emission-line fitting package QDP/PLT (Tennant 1991). For the 1997 data, the FIGARO task SDIST was used to correct for spectral curvature prior to splitting into rows, which had the undesirable effect of eliminating the variance array. 

The target properties and observation log are shown in Table 1.

\section{RESULTS}

\subsection{Extraction of extended line-fluxes}
One way to demonstrate the presence of extended line-emission is to plot the equivalent width (EW) of the detected line as a function of position along the slit. Extension then manifests itself as an off-nuclear increase in the EW if the underlying continuum originates entirely in the spatially-unresolved quasar nucleus. If, however, the continuum is extended-- and this can be ascertained by using a suitable point spread function (psf) from a standard star observation-- we use a slightly more complex method: instead of evaluating the EW with respect to the continuum measured locally in each row, we use a so-called {\em psf-scaled nuclear continuum} which is obtained by taking the continuum level in the row where it peaks (defined as the nuclear continuum), and scaling it according to the psf. The psf-scaled nuclear continuum for a given row thus represents the continuum which would have been measured in that row if there were no {\em extended} continuum. The assumption that at its peak the continuum originates entirely in the unresolved nucleus is conservative, in that where this is not the case the method underestimates the true amount of extended line flux. As stated below, Figs.~1--7 include for each object the spectrum of the [OIII] doublet integrated along the slit with the nuclear contribution overlaid, the latter being a psf-scaled version of the profile fit to the [OIII] doublet in the nuclear row (and H$\beta$ where it is within the range). Also shown are spectra of the extended [OIII] doublet alone, obtained by subtracting a psf-scaled spectrum of the nuclear row from that integrated along the slit.

The above EW procedure was used to search for extended \othree, followed by \otwo~and the narrow component of H$\alpha$. Each line complex was fitted with single component gaussian profiles atop a linear continuum, with line ratios set to their theoretical values (3:1 for both the [OIII]$\lambda5007,4959$ and [NII]$\lambda6584,6548$ doublets) and with the velocity widths of all (narrow) lines within a complex set equal. Except where the continuum was spatially extended, the broad H$\alpha$ width and the ratio of the broad H$\alpha$ flux to the local continuum were frozen at the values determined from the fit to the nuclear spectrum. Where nuclear H$\beta$ was within the wavelength range and clearly detected, it was fitted with broad and narrow components having velocity widths equal to those extracted from the H$\alpha$ and [OIII] fits, respectively; the centroids of both components were constrained to be at the same wavelength, but not tied to the position of [OIII], as the reliability of the estimated wavelength scale is questionable at the end of the wavelength range, and there are no arc-lines there with which to calibrate it. There was no evidence for any extended H$\beta$. \otwo~was typically very weak and in many cases only an upper limit on its intensity could be extracted; where it was detected, the fitting of nothing more than a single gaussian was justified at this resolution and signal-to-noise, despite the true doublet nature of this line. Errors on all quantities are the result of propagating the $\Delta \chi ^{2}=1$ parameter errors from the emission-line fitting.

For each quasar, we give a brief introduction to the literature on the source and describe the results of the EW analysis, which are also illustrated in Figs.~1--7. Extended \othree~is found in six of the objects, at the significance levels given in Table 2. The extended surface brightness in this and other lines is computed and listed in Table 3, and comparison made with the \Lya~maps of H91a and L99a. 

\subsubsection*{3C9 (Q0017+154)}
\begin{figure*}
\psfig{figure=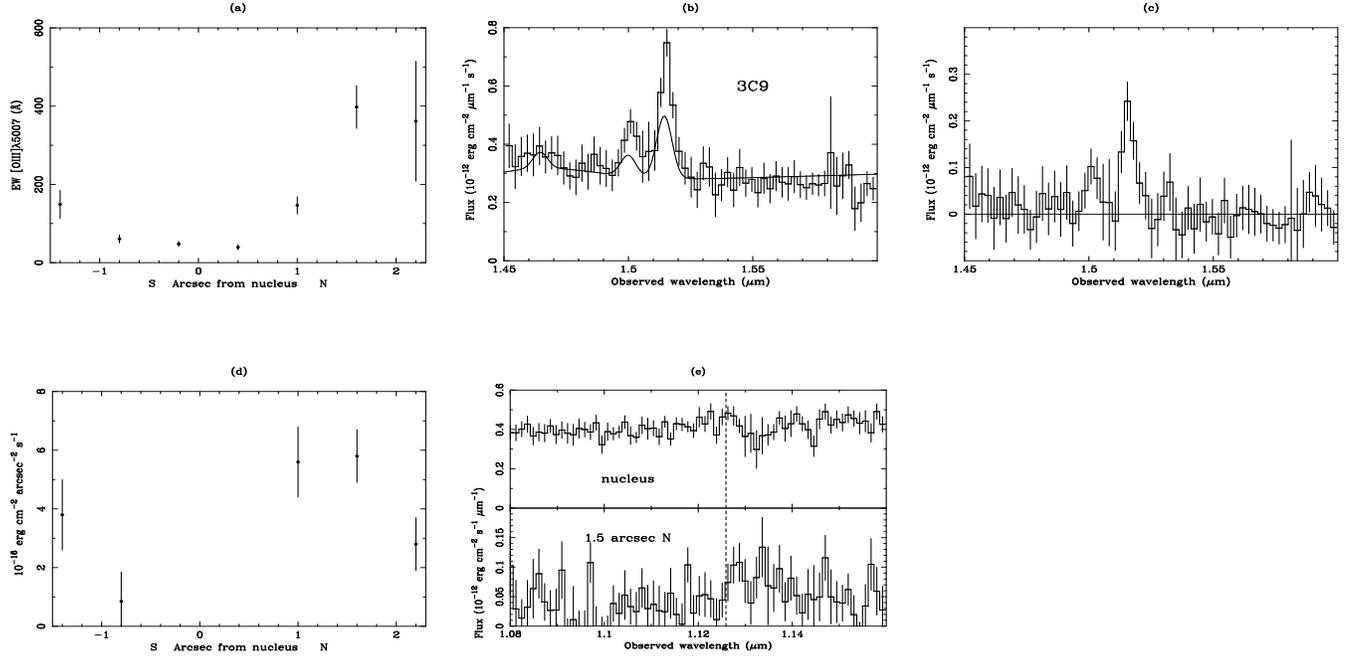,width=1.0\textwidth,angle=270}
\caption{\normalsize For 3C9: (a) EW\othree~versus nuclear distance along the N--S slit (each pixel measures 0.6~arcsec); (b) the integrated spectrum of the [OIII] doublet with the fit to the contribution of the nuclear light overlaid, the latter being the profile fit to the [OIII]+H$\beta$ complex in the nuclear row (defined as that in which the continuum peaks), scaled using the psf to include nuclear light scattered into other rows; (c) the spectrum of the {\em extended} [OIII] flux alone, obtained by subtracting from the integrated spectrum of (b) a psf-scaled spectrum of the nuclear row (the resulting line is a 10.2$\sigma$ detection); (d) extended \othree~surface brightness; (e) identified \otwo~features-- the dashed line indicates the fitted position on nucleus.}
\label{fig:3c9all}
\end{figure*}

H91a found an oval-shaped \Lya~nebula extending $\simeq 8$~arcsec north-south and misaligned with the radio source which lies along a position angle of 150 degrees (Bridle et al.~1994); a faint filament $\simeq 3.4$~arcsec to the south-east coincides with a radio hot-spot. We made the first detection of extended \othree~along a north-south slit in 3C9 during an earlier CGS4 run (Johnstone et al.~1993), prior to a chip-upgrade and through thin cirrus cloud; our current value for the redshift of the nuclear [OIII] ($z=2.021\pm0.0009$) accords well with our previous measurement ($z=2.019\pm0.001$). Fig.~\ref{fig:3c9all} shows that \othree~is clearly extended in the present data on both sides of the nucleus, especially to the north. Extended narrow H$\alpha$ is also detected in three rows but \otwo~is generally not seen, except for very marginal detections on nucleus ($z=2.019\pm0.003$) and 1.5~arcsec to the south of it ($z=2.025\pm0.002$) (see Fig.~\ref{fig:3c9all}). The underlying continuum is unresolved. 

H91a did not subtract the unresolved nuclear emission from their \Lya~map, so a comparison with our results is not straightforward. Nuclear \Lya~will contribute to their map at the radii of interest as they report a (spatially extended/total) \Lya~flux ratio of 0.2. Their lowest surface brightness contour is at $3.6 \times 10^{-17}$ \ergpcmsqparcsecps~approximately 4~arcsec north and south of the nucleus, rising to around $4.3 \times 10^{-16}$ \ergpcmsqparcsecps~1.8~arcsec to the north. Using the H$\alpha$ surface brightness in Table 3, we deduce that \Lya/H$\alpha <3.7$ in the extended gas at this position. Comparison with the dust-free case B prediction of \Lya/H$\alpha=12.3$ (Krolik \& McKee~1978) suggests that there is at least some dust present in the nebula. We note, however, that very little is needed to quench \Lya~due to the effects of resonant transfer in a dust/gas mixture (Fall et al.~1989). Indeed, since [OIII]/H$\alpha \simeq 4$ around 1.4~arcsec~N of the nucleus it follows that H$\alpha$/H$\beta \simeq 0.25 \times$[OIII]/H$\beta$, so the Balmer decrement may not substantially exceed the case B value if [OIII]/H$\beta \sim 10$, as found by Boroson et al.~(1985) for the extended gas around some quasars at $z \simeq 0.2-0.5$. The upper limit that we can place on any extended H$\beta$ does not constrain the intrinsic reddening.  

\subsubsection*{PKS 0225-014}
The Ly$\alpha$ image of H91a is marginally resolved, mostly along a PA of 140 degrees, coinciding roughly with the direction in which the central component of the radio source is extended in the radio map of Barthel et al.~(1988). A pair of non-equidistant radio lobes are located at PAs of 140 and 290 degrees. 

\begin{figure*}
\psfig{figure=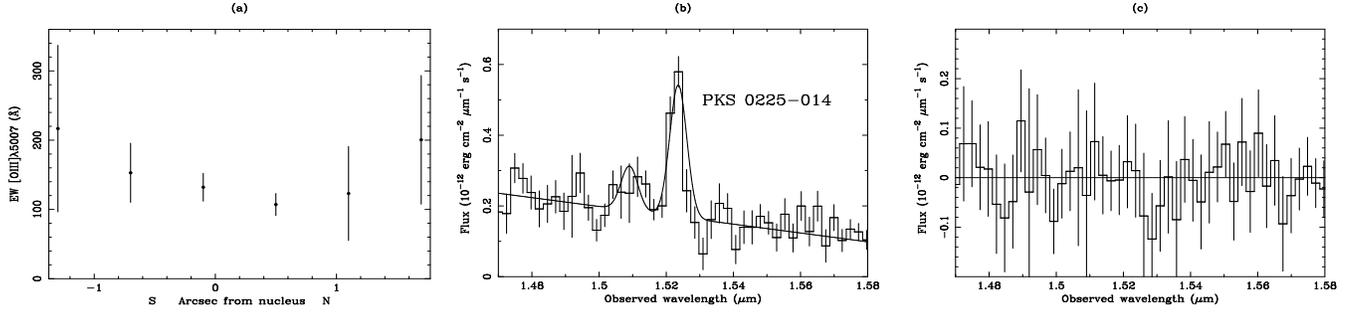,width=1.0\textwidth,angle=270}
\caption{\normalsize For PKS 0225-014: (a) EW\othree~versus nuclear distance; (b) integrated spectrum of the [OIII] doublet with the fit to the nuclear contribution overlaid; (c) extended flux spectrum.}
\label{fig:pks0225all}
\end{figure*}

We find no evidence for extended \othree~(see Fig.~\ref{fig:pks0225all}) and the underlying continuum is unresolved. There is marginal evidence for extended H$\alpha$, and comparison with the \Lya~map of H91a (from which the point source contribution has not been removed) implies that \Lya/H$\alpha < 5.6$ in the extended gas 1.2 arcsec north of the nucleus, suggesting that there is some dust associated with the gas. 

\subsubsection*{PKS 0445+097}
The Ly$\alpha$ map of H91a is asymmetrically extended to the south and south-west of the nucleus, with an overall size of 12~arcsec. Much of the radio emission orignates in a knotty jet to the WSW (Barthel et al.~1988). Lehnert et al.~(1992) reported an extended near-IR and optical continuum `fuzz', with colours similar to those of the central object (which is itself unusually red), thus ruling out the possibility that the fuzz is scattered nuclear light. The SED of the fuzz in fact resembles that of an irregular galaxy, with an inferred star formation rate of several hundred \Msunpyr. Long-slit spectroscopy by H91b found a line width of $1500 \pm 300$\kmps~FWHM in the Ly$\alpha$ nebula, with strong nuclear $z_{\rm{abs}} \simeq z_{\rm{em}}$ Ly$\alpha$ and NV$\lambda1240$ absorption. Their claim that the strong, narrow HeII$\lambda1640$ arises from the inner part of the nebula would, they say, be supported by the presence of strong [OII]$\lambda3727$ and [OIII]$\lambda5007$ with widths $>1000$\kmps. We do indeed see relatively broad \othree~both on and off-nucleus (see section~3.2). Lehnert \& Becker~(1998) found similarly broad, faint extended HeII$\lambda1640$ and CIII]$\lambda1909$ from the quasar host galaxy.

\begin{figure*}
\psfig{figure=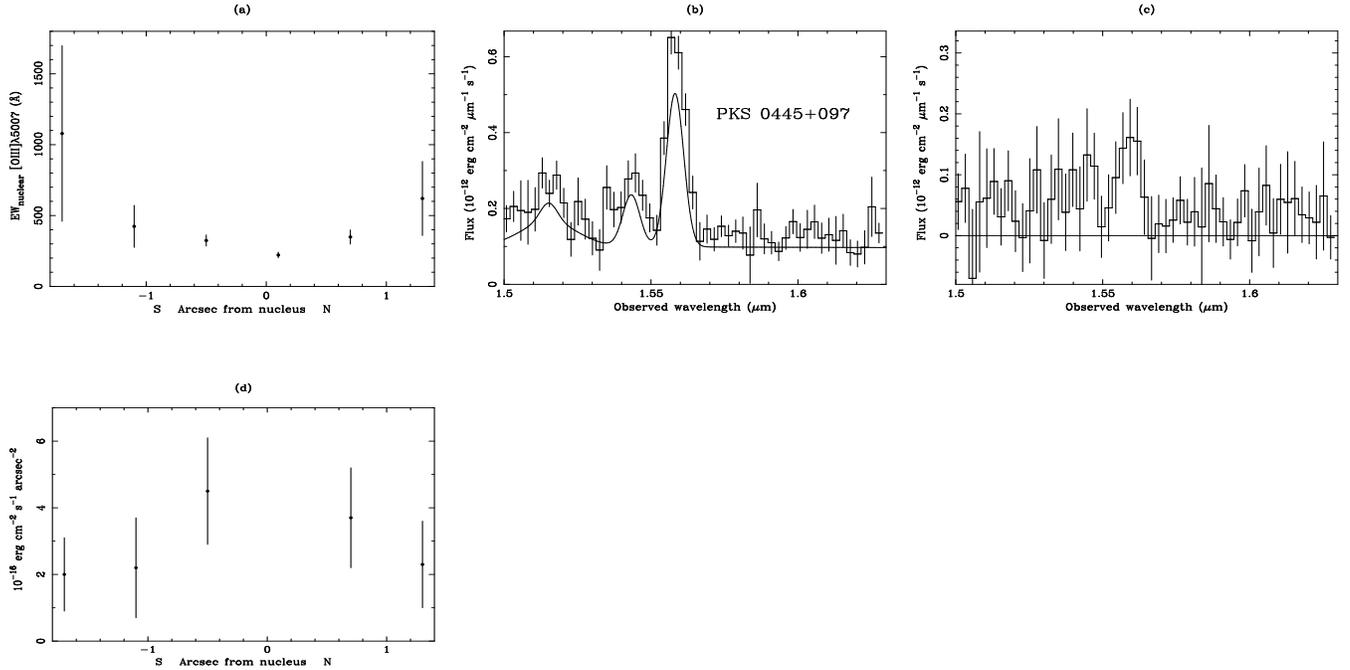,width=1.0\textwidth,angle=270}
\caption{\normalsize For PKS0445+097: (a) EW\othree~evaluated with respect to the psf-scaled nuclear continuum, because the continuum measured directly in each row is extended beyond the psf (see text for details); (b) integrated spectrum of the [OIII] doublet with the nuclear fit overlaid; (c) extended flux spectrum (the line is detected at the 6.1$\sigma$ level); (d) extended \othree~surface brightness}
\label{fig:pks0445all}
\end{figure*}

The extended continuum complicates the extended line flux computation as it is impossible to ascertain what fraction of the light is truly nuclear in origin.
As explained earlier in this section, we make the assumption that in its peak row, the continuum is entirely nuclear in origin. The EW\othree~points plotted in Fig.~\ref{fig:pks0445all} are thus calculated with respect to the {\em psf-scaled nuclear continuum}, instead of that measured locally in each row. Owing to a flux calibration problem with the J-band exposure, upper limits on EW\otwo~were converted to surface brightness limits by extrapolating the continuum from the H+K band spectra. 

The \othree~and H$\alpha$ are more extended on the southern side of the nucleus, consistent with the appearance of the \Lya~map in H91a. Using the latter, we deduce upper limits on the \Lya/H$\alpha$ ratio of 0.6 and 0.8 at distances of 0.9~arcsec north and 1.6~arscec south of the nucleus, respectively. These values are more than an order of magnitude below the case B prediction, implying that there is some dust associated with the gas.

\subsubsection*{3C191 (Q0802+103)}
Based on a slightly shorter CGS4 exposure, Jackson \& Rawlings~(1997) reported tentative evidence for extended \othree~in this object. The radio source is a linear structure along PA $10^{\circ}$ with a total extent of $\sim 5$~arcsec (Lonsdale et al.~1993) and a jet-like morphology to the south. The object is not in the H91a sample.  

\begin{figure*}
\psfig{figure=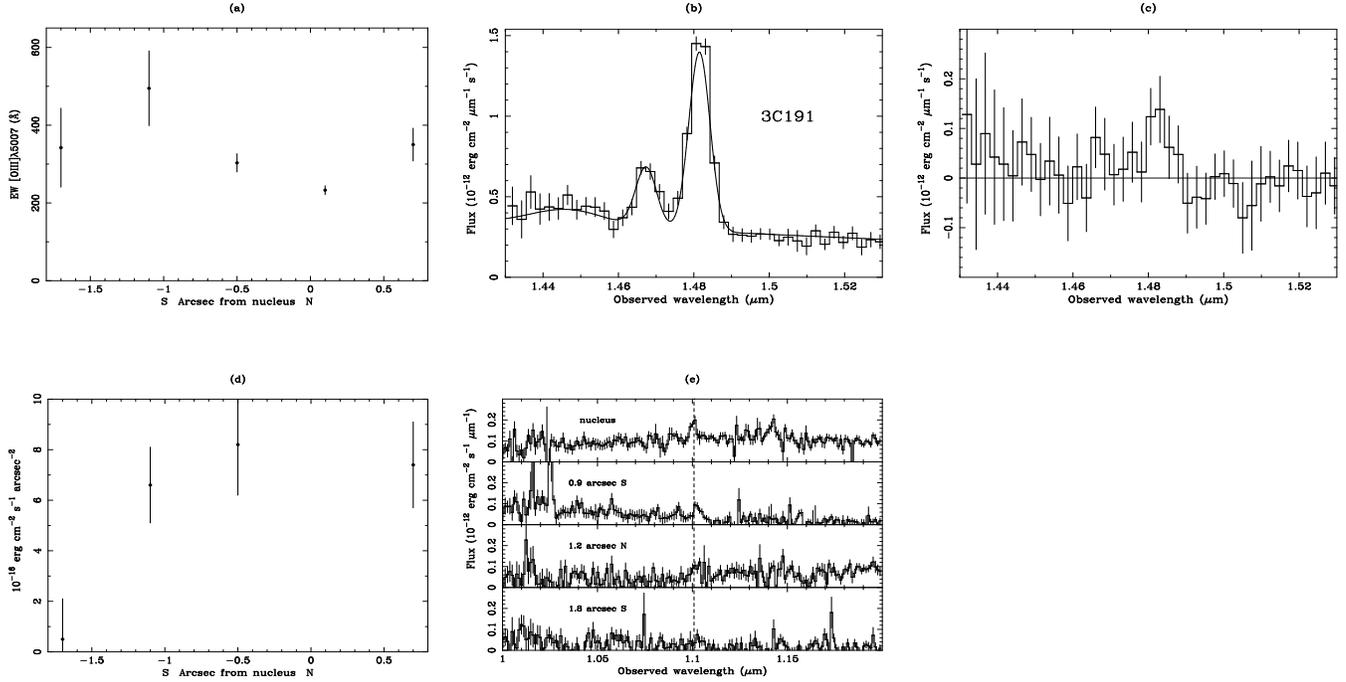,width=1.0\textwidth,angle=270}
\caption{\normalsize For 3C191: (a) EW\othree~versus nuclear distance; (b) integrated spectrum of the [OIII] doublet showing the fit to the nuclear contribution; (c) extended [OIII] spectrum (a 6.7$\sigma$ detection); (d) extended \othree~surface brightness; (e) spectra of rows with identified \otwo.}
\label{fig:3c191all}
\end{figure*}

The EW plot of Fig.~\ref{fig:3c191all} confirms that \othree~is extended, especially to the south of the nucleus. \otwo~is seen in the nuclear row and extended in three further rows, the spectra of which are also shown in Fig.~\ref{fig:3c191all}. 

\subsubsection*{4C57.29 (Q1658+575)}
\begin{figure*}
\psfig{figure=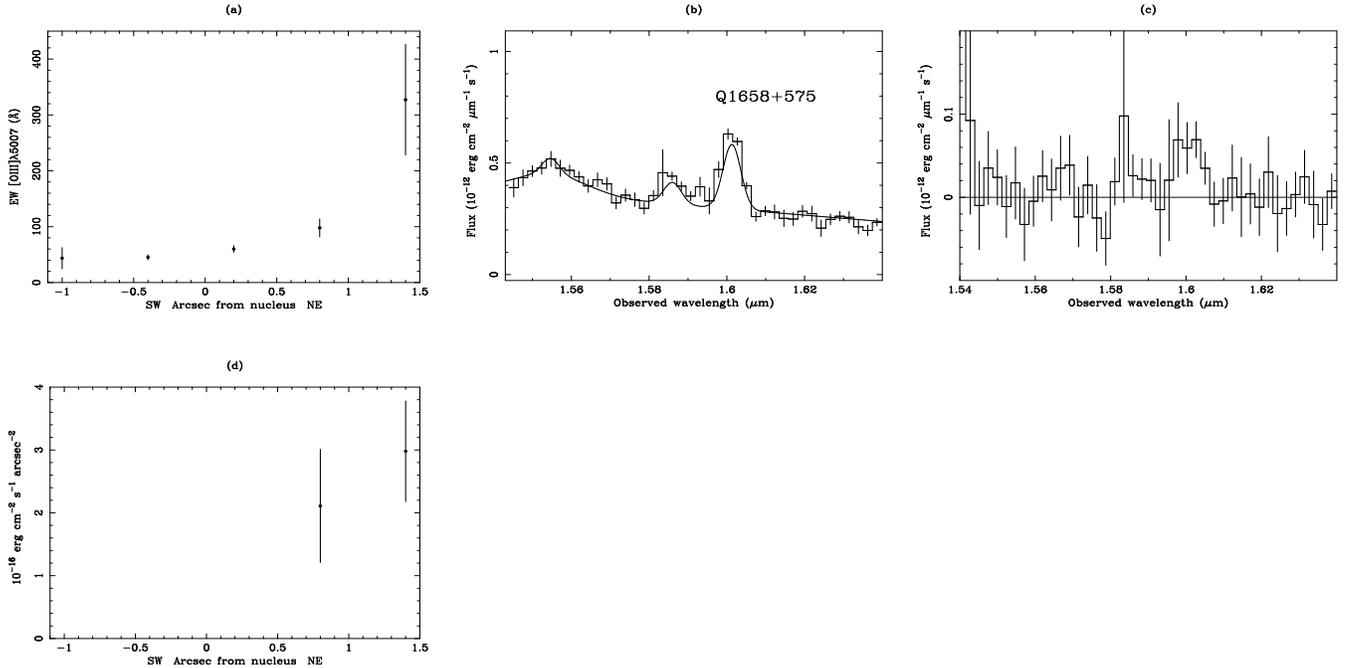,width=1.0\textwidth,angle=270}
\caption{\normalsize For Q1658+575 at a position angle of 45 degrees; (a) EW \othree; (b) integrated [OIII] doublet spectrum with the fit to the nuclear contribution overlaid; (c) extended [OIII] spectrum (a $5.4\sigma$ detection); (d) extended \othree~surface brightness.}
\label{fig:1658all}
\end{figure*}

The \Lya~map of H91a (from which the point-source {\em has} been subtracted) reveals a round nebula which is perhaps more extended to the north north-east, towards the linear radio structure along PA $18^{\circ}$ (Barthel et al.~1988). The HST narrow-band \Lya~image of L99a shows a broad swathe of emission extending 1.5~arcsec~north-east of the nucleus, as well as significant extension to the north-west, thus motivating our choice of slit position angles. Fig.~\ref{fig:1658all} corroborates these results by showing extended \othree~to the north-east. We see no evidence for extension at the other position angle but this is perhaps due to the much shorter integration time and the less-efficient 1-pixel wide slit which was in use at the start of the first night. 

For comparison, the lowest surface brightness contour in the extended \Lya~map of H91a is at $1.1 \times 10^{-17}$ \ergpcmsqparcsecps~at a radius of about 4~arcsec~in the eastern hemisphere. We thus estimate that \Lya/H$\alpha \sim 4$ in the extended gas around 1~arcsec~NE of the nucleus. Comparison with the dust-free case B prediction that \Lya/H$\alpha=12.3$ suggests that there is some dust present.  

\subsubsection*{4C47.48 (Q1816+475)}
This quasar is not in the H91a sample, but we included it on account of its extended radio structure (Barthel et al.~1988), complete with depolarization asymmetry between the jet and counter-jet sides (Garrington, Conway \& Leahy~1991). We acquired H+K band spectra along and perpendicular to the radio axis (PA 146 and 56~degrees, respectively) and also along PA 45 degrees. There was only time to do a single PA in \otwo. 

\begin{figure*}
\psfig{figure=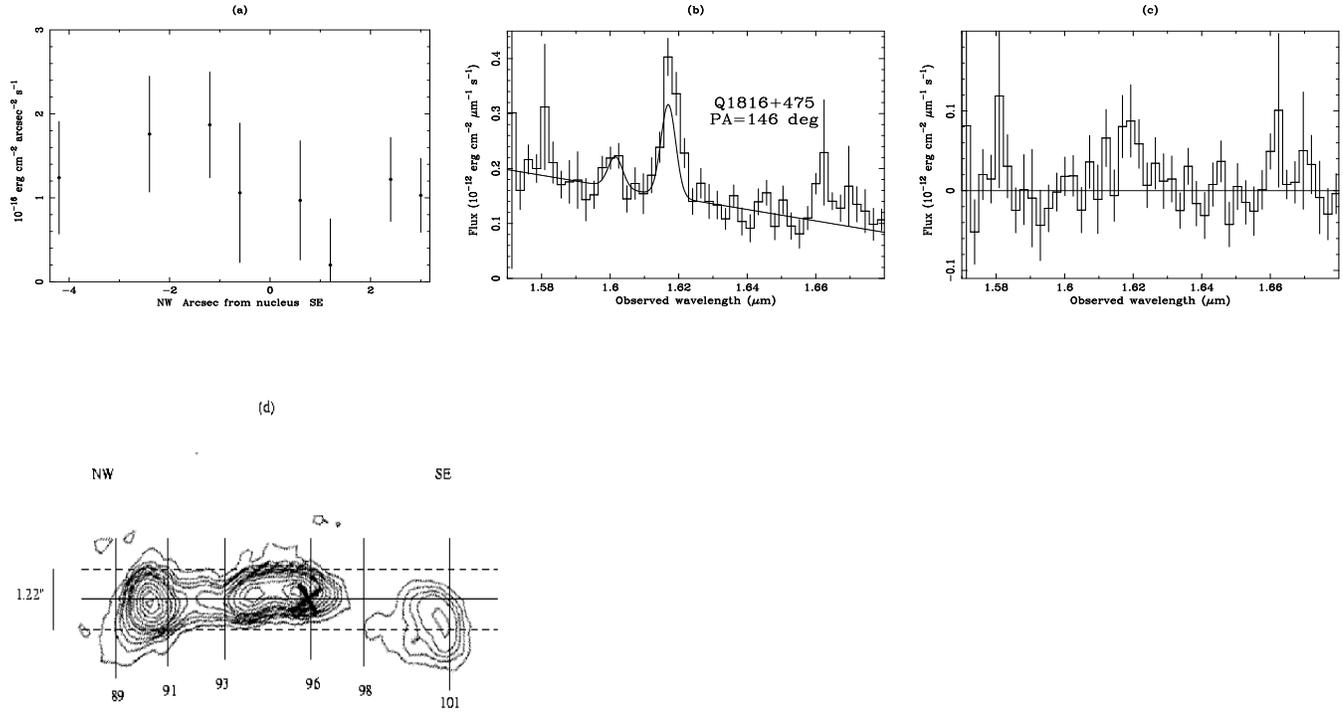,width=1.0\textwidth,angle=270}
\caption{\normalsize For Q1816+475 along PA 146 degrees: the extended \othree~surface brightness; (b) the integrated [OIII] doublet spectrum with the fit to the nuclear contribution overlaid; (c) extended [OIII] spectrum (a 4.6$\sigma$ detection); (d) VLA 6cm radio map from Barthel et al.~(1988); the grid and row numbers denote the position of the CGS4 slit (adjacent rows are 0.6~arcsec apart) and the cross marks the position of the optical nucleus according to the latter authors}
\label{fig:1816all}
\end{figure*}

Extended \othree~and H$\alpha$ are seen at both PA 146 and 56 degrees, with total [OIII] fluxes of $7.0\pm3.0$ and $1.9\pm0.8 \times 10^{-16}$\ergpcmsqps, respectively (ie. it is brighter along the radio axis). Since the error on the fitted continuum level completely dominates the uncertainty in the EW in some of the rows at large nuclear distances, we do not plot it in Fig.~\ref{fig:1816all}. For comparison with the extended \othree~surface brightness plot, we show a VLA 6cm radio map of the object, reproduced from Barthel et al.~(1988): note that [OIII] is present at the position of the fainter, south-eastern lobe (rows 100 and 101), but absent from that on the north-western side (rows 90 and 91). There is no extended \othree~along PA 45 degrees nor \otwo~of any kind.

\subsubsection*{PKS 2338+042}
\begin{figure*}
\psfig{figure=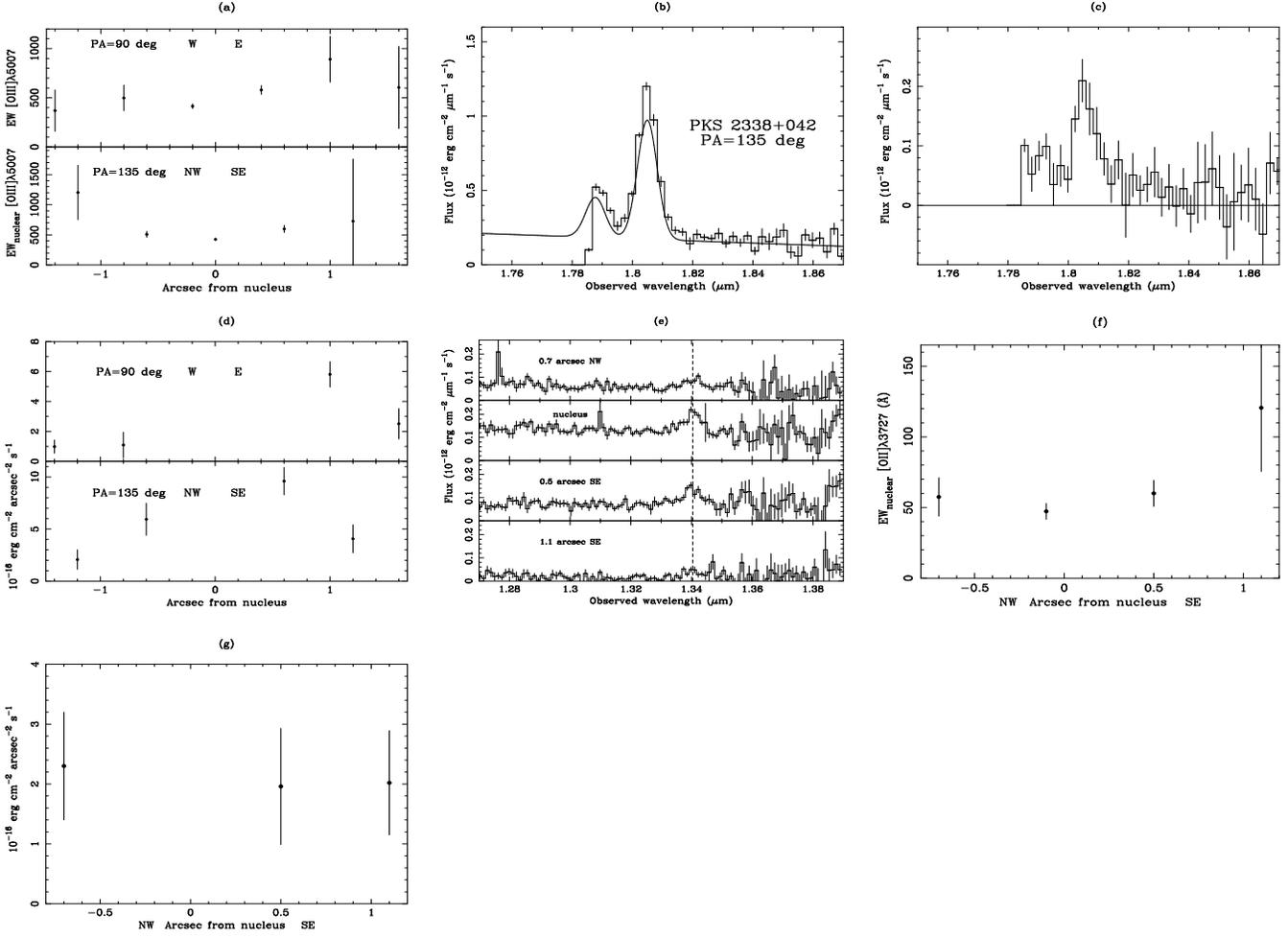,width=1.0\textwidth,angle=270}
\caption{\normalsize For PKS2338+042: (a) EW \othree~for slit position angles of 90 and 135 degrees, in the latter case evaluated with respect to the psf-scaled nuclear continuum; (b) the integrated spectrum of the [OIII] doublet with the fit to the nuclear contribution overlaid; (c) extended [OIII] spectrum at position angle 135 degrees (an 8.8$\sigma$ detection); (d) surface brightness of extended \othree~at the two position angles; (e) J-band spectra of the individual rows where \otwo~is identified (the dashed vertical line marks the nuclear wavelength); (f) EW\otwo~along PA 135 degrees, evaluated with respect to the psf-scaled nuclear continuum; (g) surface brightness of extended \otwo.}
\label{fig:pks2338all}
\end{figure*}

The \Lya~map of H91a has a north-west/south-east orientation, with brighter emission on the south-eastern side. The radio source is a smooth bent triple with lobes to the west and south-east within 2~arcsec~of the nucleus (Barthel et al.~1988). L99a interpret the spatial relationship between this distorted structure and the extended \Lya~in their HST narrow-band image as evidence for a `jet-cloud' interaction. The EWs of \othree~along the two position angles listed in Table 3 are shown in Fig.~\ref{fig:pks2338all}. That along PA 135 is computed with respect to the {\em psf-scaled nuclear} continuum, owing to the presence of extended continuum at this PA (both H91a and L99a report that the rest-frame UV is well-resolved). There is clear evidence for extended \othree~and H$\alpha$, especially to the north-west, south-east and east of the nucleus. There is also strong evidence for extended \otwo, as the EW plot of Fig.~\ref{fig:pks2338all} demonstrates for PA135 degrees (the continuum beneath \otwo~is also extended). Individual spectra of the four rows with \otwo~detections are also shown in Fig.~\ref{fig:pks2338all}. 

Comparison with the \Lya~map of H91a yields upper limits on the extended \Lya/H$\alpha$ ratio of 7, 0.8, 0.7 and 1.3 at distances of 0.9~arcsec west, 1.5~arcsec E, 1.4~arcsec north west and 1.0~arcsec south-east of the quasar, respectively, suggesting that there is some dust present.

\clearpage

\begin{table*}
\caption{Statistical significance of the extended \othree~detections}
\begin{tabular}{|ll|} \hline
Object & $\dagger$ \\ \hline
3C9           & 10.2$\sigma$  \\
PKS 0225-014  & no detection  \\
PKS 0445+097  & 6.1$\sigma$  \\
3C191         & 6.7$\sigma$  \\
Q1658+575     &  5.4$\sigma$ \\ 
Q1816+475 (PA 146 deg)    &  4.6$\sigma$  \\ 
PKS 2338+042  (PA 135 deg) & 8.8$\sigma$  \\ \hline
\end{tabular}

$\dagger$ The \othree~detection significance (in units of the standard deviation on the underlying continuum in a 4 pixel-wide extraction box) in the spatially-integrated spectrum after subtraction of a psf-scaled nuclear contribution.
\end{table*}

\begin{table*}
\begin{center}
\caption{Extended emission line surface brightnesses in units of $10^{-16}$\ergpcmsqparcsecps; a figure in the column (a,b) denotes the value in a 0.6~arcsec-wide pixel lying at some distance r from the quasar nucleus where $a \leq r<b$ (values in succesive columns {\em of a given row} correspond to points separated by the 0.6 arcsec pixel size); the tabulated values should {\em not} be understood as average surface brightness levels within the stated range. The slit orientations are as indicated.}

\begin{tabular}{|rlllllllll|} \hline 
    &    \multicolumn{9}{c}{Nuclear distance along the slit in arcsec} \\
Object    &(-2.4,-1.8) & (-1.8,-1.2)    & (-1.2,-0.6) & (-0.6,0)     & (0,0.6) & (0.6,1.2)   & (1.2,1.8)   & (1.8,2.4) & $\left|r\right|\geq2.4$\\ \hline
3C9:       & (S)   &               &             &              &          &             &            &     & (N)       \\ 
\rm{[OIII]} & --  & $3.8 \pm 1.2$ & $0.9\pm1.0$ & --       & --    & $5.6\pm1.2$ & $5.8\pm0.9$ & $2.8\pm0.9$ & -- \\
H$\alpha$ & --         & $1.1\pm0.8$    & --0.8$\pm$0.9 & -- & $0.08\pm0.9$ & 1.4$\pm0.7$ & 1.2$\pm$0.6 & --   & -- \\
\rm{[OII]} & --     & $<7.3$    & $<8.6$      & --           & --      & $<8.5$      & 3.8$\pm1.2$  & $<8.9$ & -- \\ \\
PKS0225-014: & ~\rm{(S)}  &               &             &              &          &             &       & & ~\rm{(N)}       \\ 
H$\alpha$ & --       & --             & -- & 2.1$\pm$1.3      & 0.8$\pm$1.2& 0.7$\pm$1.1 & --          & -- & -- \\ \\
PKS0445+097: &   ~\rm{(S)}  &               &             &              &          &             & &            & \rm{(N)}      \\ 
\rm{[OIII]} & -- & 2.0$\pm$1.1 & 2.2$\pm$1.5 & 4.5$\pm$1.6 & -- & 3.7$\pm$1.5 & 2.3$\pm$1.3 & -- & -- \\ 
H$\alpha$ & --       & 1.6$\pm$1.1  & 1.9$\pm$1.2  & 5.2$\pm$2.6 & -- & 2.7$\pm$2.3 & -- & -- & -- \\ 
\rm{[OII]} & --     & $<2.0$ & $<2.3$ & $<2.8$  & $<2.4$  & $<2.0$  & --  &  --  & -- \\ \\
3C191:& \rm{(S)}    &               &             &              &          &      &      &      &  \rm{(N)}     \\ 
\rm{[OIII]} & -- & 0.5$\pm$1.6 & 6.6$\pm$1.5 & 8.2$\pm$2.0 & -- & 7.4$\pm$1.7 & --0.9$\pm$1.3 & -- & -- \\ 
H$\alpha$ & --   & 1.9$\pm1.0$ & 2.6$\pm2.0$ & $3.0\pm3.2$ & -- & --0.2$\pm$2.0 & 2.3$\pm$1.1 & -- & -- \\ 
\rm{[OII]} & 0.8$\pm$0.6 & -- & 1.4$\pm$0.7 & -- & -- & 1.2$\pm$0.7 & -- & -- & -- \\ \\
Q1658+575: & \rm{(SW)}   &               &             &              &          &             &            &      & \rm{(NE)}      \\ 
\rm{[OIII]} & -- & -- & --0.2$\pm$0.7& -- & -- & 2.1$\pm$0.9 & 3.0$\pm$0.7& -- & -- \\
H$\alpha$ & --  & -- & --0.04$\pm$0.3 & 0.6$\pm$0.6 & -- & 0.3$\pm$0.5 & 0.5$\pm0.3$ & -- & -- \\ \\
Q1816+475: & ($\sim$\rm{NW})  &               &             &   &          &             &            &            & ($\sim$\rm{SE})     \\
\rm{[OIII]} &1.8$\pm$0.7& -- & 1.9$\pm$0.6& 1.1$\pm$0.8& -- & 1.0$\pm$0.7 & 0.2$\pm$0.5 & -- & $\dagger$ \\
H$\alpha$ & --  & 0.4$\pm$0.2 & 0.2$\pm$0.3 & -- & 0.3$\pm$0.4 & -- & 0.4$\pm$0.2 & -- & -- \\
Q1816+475: & ($\sim$\rm{SW}) &               &             &              &          &     &      &    & ($\sim$\rm{NE})    \\
\rm{[OIII]} & -- & -- & -- & --0.5$\pm$0.8 & -- & 0.7$\pm$0.7 & 0.9$\pm$0.6 & 0.9$\pm$0.5 & -- \\
H$\alpha$ & --  & 0.3$\pm$0.3 & --0.5$\pm$0.4 & -- & 0.6$\pm$0.6 & 0.1$\pm$0.3 & 0.6$\pm$0.4 & -- & -- \\ \\
PKS2338+042:& \rm{(NW)}  &               &             &              &          &    &            &      & \rm{(SE)}       \\
\rm{[OIII]} & -- & -- & 2.1$\pm$0.9 & 5.9$\pm$1.5 & -- & 9.6$\pm$1.3 & 4.1$\pm$1.3 & -- & --\\
H$\alpha$ & --  & 1.6$\pm$0.7 & 6.6$\pm$1.4 & -- & -- & 3.8$\pm$1.9 & -- & -- & -- \\
\rm{[OII]} & --  & $<2.90$   & 2.3$\pm$0.9 & -- & 2.0$\pm$1.0 & 2.0$\pm$0.9 & $<4.3$ &  --  & -- \\ 
PKS2338+042:& \rm{(W)}  &        &      &     &    &       &            &            & \rm{(E)}       \\
\rm{[OIII]} & -- & 1.0$\pm$0.4& 1.1$\pm$0.9 & -- & -- & 5.8$\pm$0.9 & 2.5$\pm$1.0& -- & --\\
H$\alpha$ & --  & -- & 1.0$\pm$0.8 & -- & -- & 9.3$\pm$1.4 & 3.1$\pm$1.2 & -- & -- \\ \hline

\end{tabular}
\end{center}
$\dagger$ Q1816+475 shows extended \othree~at $\left|r\right|\geq2.4$ arcsec along the radio axis (PA 146 degrees), with values of $1.2\pm0.7$, $1.2\pm0.5$ and $1.0\pm0.4 \times 10^{-16}$\ergpcmsqparcsecps~at $r=-4.2$, $+2.4$ and $+3.0$ arcsec from the nucleus, respectively.
\end{table*}

\clearpage

\clearpage

\clearpage

\clearpage

\section{DISCUSSION}

Before considering the detailed properties of the extended line emission, we mention the rather obvious point that the mere fact that it is seen from a coolant such as [OIII] implies that the galactic-scale environment of the quasar has undergone some metal-enrichment. An episode of star-formation has thus occurred, either concurrently with or prior to the quasar phase itself. This is also suggested by the presence of dust associated with the gas, as evidenced by \Lya/H$\alpha$ ratios well below the case B value.

\subsection{Kinematics and morphology of the extended line-emission}
The noise level of the present data do not permit us to examine the kinematic structure of the gas or its morphological association with the extended radio structure in spatially-resolved detail. In fact, nothing more than the extraction of a single, integrated spectrum of the off-nuclear light can really be justified for kinematic purposes. We thus fitted a single kinematic component to the [OIII] doublet of each of the off-nuclear spectra, to determine an average velocity offset (from the nuclear emission) and the velocity dispersion. The results are shown in Table 4.

Whilst the errors for individual objects are large, the extended gas is on average redshifted by a few hundred \kmps~relative to the nuclear emission. Furthermore, although the errors are larger still, we found a velocity offset of the same sign on both sides of the nucleus within a given object. This tendency for the extended [OIII] to be redshifted from the nuclear [OIII] can be seen in Figs.~1-7, where we show for each quasar an integrated spectrum of the [OIII] doublet with the fit to the spatially-unresolved nuclear component overlaid: residual emission tends to appear redward of the fit. One possible explanation for this lack of kinematic continuity is that the nuclear [OIII] is blueshifted with respect to the systemic redshift of the galaxy at which the bulk of the extended emission arises. 

\begin{figure}
\psfig{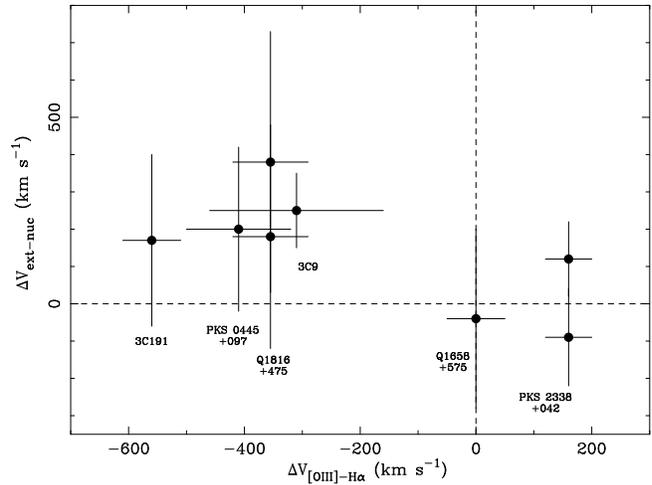}
\caption{\normalsize The extended--nuclear \othree~velocity difference ($\Delta V_{\rm{ext-nuc}}$) plotted against the nuclear \othree--H$\alpha$ velocity difference ($\Delta V_{\rm{[OIII]-H\alpha}}$).}
\label{fig:vshifts}
\end{figure}

In this connection, there have been several studies of the velocity shifts between quasar nuclear emission lines of different ionization; using a sample of 160 quasars, Tytler \& Fan~(1992) found that each ultraviolet emission line has a well-defined mean velocity with respect to a systemic value defined by H$\beta$, increasing roughly in the order of increasing ionization to $-454\pm37$\kmps~for HeII$\lambda1640$. They also confirmed the result of Gaskell~(1982) that in low-z quasars and Seyfert galaxies, the Balmer and narrow line redshifts agree with one another to within 100\kmps. Wilkes~(1986), however, found that the [OIII]$\lambda5007,4959$ lines were blueshifted by a mean velocity of $220\pm150$\kmps~from H$\beta$. Applied to our sample, the latter result could account for the velocity offsets between the extended and nuclear \othree. To investigate this hypothesis further, we measured the velocity of nuclear \othree~ relative to H$\alpha$, for comparison with the $\Delta V_{\rm{ext-nuc}}$ values in Table~4, as shown in Fig~\ref{fig:vshifts}. Despite the large errors, the sense of the apparent trend is consistent with an interpretation in which the nuclear \othree~is blueshifted from both the extended \othree~and the nuclear H$\alpha$ by comparable amounts.

\begin{table}
\caption{Kinematic properties of the off-nuclear gas}
\begin{tabular}{||llll||} \hline
Object & PA    & $\Delta V_{\rm{ext-nuc}}\star$ & FWHM         \\  
       & (deg) & (\kmps)           &  (\kmps)     \\  \hline          
3C9 & 0 & $250 \pm 100$ & $520^{+320}_{-520}$\\ 
PKS 0225-014 & 0 & $\dagger$ & $\dagger$ \\
PKS 0445+097 & 0 & $200 \pm 220$ & $1030^{+520}_{-730}$\\
3C191 & 0 & $170 \pm 230$ & $610^{+970}_{-610}$   \\
Q1658+575 & 45 & $-40 \pm 240$ & $1120^{+540}_{-680}$\\
Q1816+475 & 146 & $380 \pm 350$ & $1260^{+1300}_{-1260}$\\
	  &  56 & $180 \pm 300$ & unresolved \\
PKS 2338+042 & 90 & $-90 \pm 130$ & $1180^{+390}_{-430}$ \\ 
                       & 135 & $120 \pm 100$ & $980^{+270}_{-300}$ \\ \hline
\end{tabular}

$\star$ Velocity of the extended [OIII] emission with respect to that in the nucleus. \\ 
$\dagger$ The extended flux is not visible above the noise.
\end{table}

\begin{figure*}
\psfig{figure=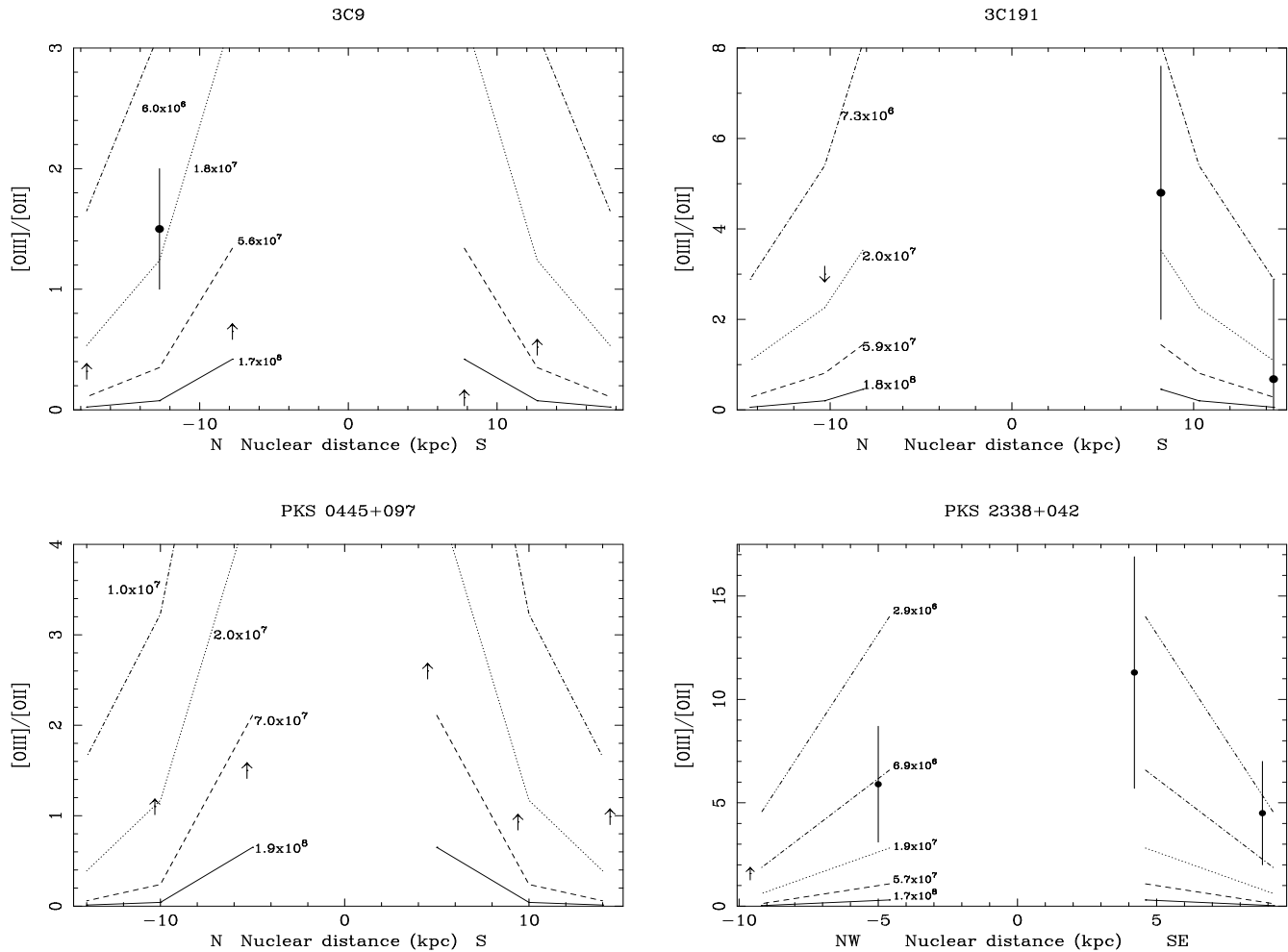,width=1.0\textwidth,angle=270}
\caption{\normalsize Radial variation along the slit of the [OIII]/[OII] ratio in the four quasars for which we have such information, together with the predictions of constant pressure CLOUDY photoionization models which are labelled with the values of the total gas pressure to which they correspond (in units of \pcmK).}
\label{fig:pressure}
\end{figure*}

Alternatively, if the nuclear \othree~is produced at the systemic velocity, the tendency of the extended emission to be redshifted relative to it could instead be due to gaseous {\em infall} towards the quasar. In this model, dust between the line-emitting clouds would be invoked to obscure those on the far side which would otherwise appear blueshifted. Evidence for infall is provided by the large line widths in the extended gas-- our values are compatible with those measured by H91b for the off-nuclear \Lya~(1000--1500\kmps FWHM) in some of these quasars and in others near this redshift. Such large line widths, they argued, result either from infall into the potential well of a very massive galaxy or from interaction with the radio-emitting plasma, but they did not find that the more kinematically-disturbed regions of \Lya~emission were associated with the strongest radio emission. As we noted earlier, however, L99a found evidence from HST data for a `jet-cloud' interaction in PKS 2338+042: they demonstrated that the cloud responsible for the \Lya~emission is massive enough to deflect the radio jet and to survive for long enough after the collision to emit at the observed epoch, several Myr later. Our present data provide little new information on the detailed relation of the extended line-emission to the radio structures. The 1997 observations were at a single position angle and the line emission in two of the 1999 objects, Q1658+575 and PKS 2338+042, was already known to align more closely with the radio structure than perpendicular to it (H91a; L99a). Our results on Q1816+475 are qualitatively similar, with the strongest extended emission being along the linear radio axis, as remarked above. Concerning dust, we have noted that while the \Lya/H$\alpha$ values in the extended gas are typically well below the case B dust-free prediction, this does not necessarily indicate substantial amounts of reddening. Where H$\beta$ is within the wavelength range, upper limits on its extended flux do not place any constraints on the extinction.

On a global level, where the slit is aligned in the general direction of the extended radio structure, the extended [OIII] tends to be brighter on the side of the nucleus with the stronger, jet-like, radio emission, as in PKS 0445+097, 3C191, Q1816+475, Q1658+575 and PKS 2338+042. The only exceptions to this are 3C9 and PKS 0225-014 (in the latter we see no extended \othree~along a north-south slit position). H91a reported that such a shared asymmetry also exists in the extended radio and \Lya~emission, but owing to the greater extinction suffered by \Lya~on the far side of the nebula, they could not determine whether this was genuinely due to the line-emitting gas being distributed around the quasar in an intrinsically anisotropic fashion. That the less-dust-sensitive [OIII] and H$\alpha$ emission are at least as asymmetric as the \Lya~strongly suggests that it is.

As stated by H91a, an intrinsically anisotropic gas distribution is in line with theoretical models for galaxy formation (eg. Rees~1988), as the Hubble time at this redshift would scarcely have been long enough to permit the establishment of a symmetric gas configuration in such a massive halo. Models in which quasar activity is triggered by galaxy interactions (eg. Calberg~1990) also predict anisotropic gas distributions. McCarthy et al.~(1991) found a similar asymmetry in powerful radio galaxies, with the extended optical line emission virtually always being brighter on the side of the closer radio lobe. There are hints, in eg. Q1816+575 in Fig.~6, of a more detailed correspondence between the extended [OIII] and the radio plasma, but an indepth astrophysical analysis thereof requires near-IR spectroscopy on 8m-class telescopes. Such correlations could, however, be accounted for directly if the line-emission were powered by shocks associated with the radio source, as found by Best et al.~(2000a,b). 

\subsection{The [OIII]/[OII] ratio and the extended gas pressure}
A key objective of this work was to measure the [OIII]/[OII] ratio in the extended gas as a diagnostic of the pressure, for comparison with the lower redshift studies listed in section~1. As mentioned in these papers, the frequent occurrence of high pressure line-emitting gas around radio-loud objects requires that the warm clouds be long-lived; this in turn suggests that they are pressure confined by some hotter medium (lest they would disperse rapidly on a sound-crossing time), an obvious candidate being a medium similar to the hot X-ray-emitting gas found in clusters of galaxies (the ICM). In the light of the findings of subsection~4.1, it is likely that in those objects where the slit is aligned essentially along the radio axis (ie. 3C191, Q1816+475 and PKS 2338+042) that the line emitting clouds are compressed by passage through shocks associated with the radio jets and that the pressures so-calculated are not representative of conditions in the surrounding medium (see eg. Best et al.~2000b). To compute the pressure, we adopt a model in which ionization-bounded (optically-thick) clouds are subject to the photoionizing continuum of the central quasar, so the neglect of any in situ sources of ionization (eg. shocks from the radio jet) results in the derived pressures being lower limits on the actual values. A detailed discussion of the reliability of these assumptions may be found in Bremer~(1993). We note that the derived pressures would increase if an extinction correction were applied and decrease if the true nuclear distance to the clouds were used, as opposed to the projected value.

We have information on the [OIII]/[OII] ratio in the extended gas for four of the seven quasars, although for one of them-- PKS 0445+097-- it is wholly in the form of lower limits on the line ratio. The measured [OIII]/[OII] values are also subject to uncertainty, for the [OIII] and [OII] fluxes are derived from separate long-slit spectra, which may not sample the same part of the object and between which there may be flux calibration differences. To investigate the latter we have compared the observed J and H band nuclear continuua for each quasar, and found that for each of 3C9 and 3C191, the one matches well the extrapolation of the other; for PKS 2338+042, however, the `raw' [OIII]/[OII] values have been increased by a factor of 2.25 as the J-band continuum exceeds the extrapolation by this factor (which is not surprising since the ratioing star used for the J-band exposure was not flux-calibrated against a faint standard). Recall that the J-band spectrum of PKS 0445+097 could not be flux calibrated in the usual manner. We are not able to perform a correction for intrinsic dust extinction.

Version 90.05 of the photoionization code CLOUDY was employed (Ferland~1996), using the `AGN' model continuum therein, which comprises an exponentially cut-off `Big Bump' component and an X-ray power law:

\begin{equation}
f_{\nu} = \nu^{\alpha_{uv}}exp(-h\nu/kT_{BB})exp(-kT_{IR}/h\nu) + a\nu^{\alpha_{x}}
\end{equation}

The coefficient $a$ is adjusted by the code to produce the correct X-ray (2\keV) to optical (2500\AA) spectral index, $\alpha_{\rm{ox}}$, for the case where the Big Bump does not contribute to the emission at 2\keV. The default parameter values, representative of a `typical AGN', are $T_{BB}=1.5\times 10^{5}$K, $\alpha_{\rm{ox}}=-1.4$, $\alpha_{\rm{uv}}=-0.5$, $\alpha_{\rm{x}}=-1$~and $kT_{IR}=0.01$Ryd. The X-ray power law is added only for energies greater than 0.1~Ryd, and above 100\keV~the continuum is assumed to fall off as $\nu^{-3}$. Where available, literature information on the quasars was used to adjust the parameters $T_{BB}$, $\alpha_{\rm{ox}}$, $\alpha_{\rm{uv}}$ and $\alpha_{\rm{x}}$ to reconstruct the observed portions of the SEDs. The parameters were otherwise left at their default values. For 3C9 and 3C191, we take $\alpha_{\rm{ox}}$ values (of --1.478 and --1.405, respectively) and X-ray luminosities from Worrall et al.~(1987); for PKS 0445+097, we take U, B, R and K photometry from Lehnert et al.~(1992) and H91a, along with a {\em ROSAT} X-ray flux (Siebert et al.~1998), to obtain $T_{BB}=7.2\times 10^{4}$K, $\alpha_{\rm{ox}}=-1.0$, $\alpha_{\rm{uv}}=-1$ and $\alpha_{\rm{x}}=-1$; similarly, for PKS 2338+042 we find from B-band photometry (H91a) and an HST flux density at $\lambda_{\rm{rest}}=1520$\AA~(L99a), that this narrow portion of the SED is well-matched by the default parameter values.

For each object, we ran a grid of constant pressure CLOUDY models for clouds at various radii from the quasar with different values of the hydrogen density at the exposed cloud face. The results are shown in the plots of Fig.~\ref{fig:pressure}, where the CLOUDY model loci are labelled with the value of the gas pressure ($n_{\rm{gas}}T_{\rm{gas}}$) to which they correspond. With the exception of PKS~2338+042 (where in any case the neglect of in situ ionization associated with the jet-cloud interaction renders the derived pressure a lower limit), our pressure measurements are equal to the maximum possible values that could be maintained at these radii by an isobarically-cooling, pressure-confining ICM in an isothermal potential with a line of sight velocity dispersion $\sigma_{\rm{los}}$ in the range $300 < \sigma_{\rm{los}}< 400$\kmps (see Bremer et al.~1992 for details); the majority of the limits on the pressure are also consistent with such a high pressure environment, but do not otherwise impose strong constraints on the gas properties. Our values (and limits) are comparable to the pressures measured by Bremer et al.~(1992) at slightly larger radii ($\sim 20$\kpc) around some radio-loud quasars at redshift $z \sim 1$, and thus not obviously inconsistent with their conclusion that the most dramatic pressure evolution seen so far is the strong decrease between redshift $z \sim 1$ and the present day (see their Fig.~12).

\section{The nuclear spectra}
Although we have hitherto been concerned with the extended emission around the quasars, their nuclear spectra also merit attention in the light of recent work on radio galaxies and quasars at lower redshift. Jackson \& Rawlings (1997) found that radio galaxies and quasars at $z>1$ lie on essentially the same correlation between the [OIII] and extended 178 MHz radio luminosities, whereas at lower redshift quasars had been found to be more [OIII]-luminous; this finding has implications for quasar/radio galaxy unification schemes. Similarly, Willott et al.~(1999) used the 3C and (much fainter) 7C surveys to deduce that the [OIII] and [OII] luminosities of low redshift radio galaxies depend primarily on the radio power rather than redshift. Tadhunter et al.~(1998) reported that the [OII]/[OIII] versus $L_{\rm{[OIII]}}$ anti-correlation which exists for radio galaxies at $z<0.2$ (Saunders et al.~1989) breaks down at higher redshift ($0.5<z<0.7$), and similar conclusions were reached by Jackson \& Rawlings~(1997). Our present data give some indication of the position of higher redshift quasars on these correlations.  

The nuclear \othree~and \otwo~fluxes and [OIII]/[OII] ratios are tabulated in Table 5. When placed on Figs.~4 (nuclear $L_{\rm{[OIII]}}$ versus extended 178Mhz radio luminosity, $L_{\rm{178~MHz,ext}}$) and 7 ([OIII]/[OII] versus $L_{\rm{178~MHz,ext}}$) of Jackson \& Rawlings~(1997), our data occupy the same regions as their lower redshift quasars (after 3C191 and the radio galaxy 3C257 at $z=2.48$, their highest redshift object is the $z=1.681$ radio galaxy 3C222; the mean redshift of their quasars is $z=1.37$). $L_{\rm{178~MHz,ext}}$ is considered to be a good isotropic indicator of the power of the central engine, for radio galaxies and quasars, and was estimated for our objects using data in Garrington, Conway \& Leahy~(1991), Barthel \& Lonsdale~(1983), Barthel et al.~(1984), Drinkwater et al.~(1997), Laing et al.~(1983) and Reid et al.~(1995).

This finding is consistent with there having been little evolution in the emission-line properties of radio-loud quasars between redshifts $z \sim 2.2$ and $z \sim 1.4$. We do, however, caution that it is based on a small, statistically incomplete, sample of objects.

\begin{table}
\caption{Nuclear emission-line fluxes}
\begin{tabular}{|llll|} \hline
Object & \othree & \otwo & [OIII]/[OII] \\  \hline
3C9           &   $19.2\pm2.3$ & $7.5\pm3.4$  & $2.6\pm1.2$ \\ 
PKS 0225-014  &   $23.8\pm5.9$ & --           & --          \\
PKS 0445+097  &   $26.2\pm4.4$ & $<4.9$       & $>5.3$      \\
3C191         &   $74.4\pm5.0$ & $9.5\pm2.1$  & $7.8\pm1.8$ \\
Q1658+575     &   $16.7\pm1.2$ & --           & --          \\
Q1816+475     &   $9.7\pm2.3$  & $<1.5$       & $>6.5$      \\ 
PKS 2338+042  &   $56.7\pm9.0$ & $14.6\pm2.0$ & $8.8\pm1.8\dagger$ \\ \hline
\end{tabular}
Fluxes are in units of $10^{-16}\ergpcmsqps$; where more than one PA was observed in [OIII], the listed value is the mean of the measurements.\\  
$\dagger$ In deriving this figure, the `raw' flux ratio was increased by a factor of 2.25 to take account of the flux-calibration mismatch between the J and H-band spectra, as described in section~4.2 
\end{table}

\section{SUMMARY AND CONCLUSIONS}
To summarise, we have made significant detections of extended \othree~and H$\alpha$ line-emission around six out of a sample of 7 radio-loud quasars at $z \sim 2$, five of which were known from H91a and L99a to have \Lya~nebulae. In conjunction with the latter works, our findings lead to a number of conclusions about the gaseous environment within a projected radius of $\sim 20\kpc$ of the quasar. 

Firstly, the mere existence of extended [OIII] emission and the dust associated with it implies that the galactic-scale environment of the quasar has already been enriched by an episode of star formation.

Secondly, we found that the extended [OIII] tends to be brighter on the side of the nucleus with the stronger radio emission, indicating that the extended gas is distributed anisotropically, and perhaps in some cases also compressed and shock-ionized through interaction with the radio plasma.

Thirdly, on a global scale, the extended [OIII] is typically redshifted by a few hundred \kmps~from the nuclear [OIII]. By comparison with the measured velocity difference between the nuclear [OIII] and H$\alpha$, we believe that this is probably due to the nuclear [OIII] being blueshifted by comparable amounts from a systemic value defined approximately by both the extended emission and the nuclear H$\alpha$. We cannot, however, rule out an alternative model involving gaseous infall coupled with the effects of selective dust extinction.

Fourthly, for four of the quasars we have information on the [OIII]/[OII] ratio in the extended gas. This was used in conjunction with CLOUDY photoionization modelling to place constraints on the gas pressure, assuming that the radiation from the quasar itself is the only source of ionization. We cautioned, however, that such a scenario may not hold in all cases in view of the observed association between the radio plasma and the emission-line gas. With the exception of PKS 2338+042, we measure gas pressures (or limits thereon) which are comparable to the maximum pressure which could be maintained at these radii by any isobarically-cooling ICM. This suggests that such an ICM would host a strong cooling flow, with a mass deposition rate high enough to significantly influence the fuelling of the quasar, as well as the formation of the host galaxy (see Fabian \& Crawford~1990). Future observations with Chandra and XMM are expected to reveal such ICMs directly through their X-ray emission. The case of A3581, however, illustrates that emission-line spectroscopy may continue to provide a complementary pressure diagnostic for the environments of high-redshift quasars; X-ray observations by Johnstone et al.~(1998) of this $z\simeq0.02$ poor cluster (containing the radio galaxy PKS 1404-267) show that despite its low temperature ($\sim 1.5-2.0$\keV) and relatively low luminosity ($\sim 2 \times 10^{42}$\ergps), its mass deposition rate is as high as $\sim 80$\Msunpyr.

\section*{ACKNOWLEDGMENTS} 
UKIRT is operated by the Joint Astronomy Centre on behalf of the United Kingdom Particle Physics and Astronomy Research Council. RJW and RMJ thank PPARC for financial support, CSC the Royal Society. Kelvin Wu is thanked for assistance during the 1997 observing run. The anonymous referee is thanked for helpful comments. This research has made use of the NASA/IPAC Extragalactic Database (NED).

{}

\end{document}